\documentclass[11pt,a4paper]{article}
\usepackage{cite}
\usepackage{amsmath, amsthm, amssymb,slashed,upgreek}
\usepackage{ifpdf}
\ifpdf
  \usepackage[pdftex]{graphicx}
  \usepackage{epstopdf}
\else
  \usepackage[dvips]{graphicx}
\fi
\parskip=\baselineskip
\def\Bbb{\mathbb}

\def\16{{\bf 16}}
\def\1{{\bf 1}}
\def\2{{\bf 2}}
\def\3{{\bf 3}}
\def\4{{\bf 4}}

\def\bar{\overline}

\font\teneurm=eurm10 \font\seveneurm=eurm7 \font\fiveeurm=eurm5
\newfam\eurmfam
\textfont\eurmfam=\teneurm \scriptfont\eurmfam=\seveneurm
\scriptscriptfont\eurmfam=\fiveeurm

\font\teneusm=eusm10 \font\seveneusm=eusm7 \font\fiveeusm=eusm5
\newfam\eusmfam
\textfont\eusmfam=\teneusm \scriptfont\eusmfam=\seveneusm
\scriptscriptfont\eusmfam=\fiveeusm

\font\tencmmib=cmmib10 \skewchar\tencmmib='177
\font\sevencmmib=cmmib7 \skewchar\sevencmmib='177
\font\fivecmmib=cmmib5 \skewchar\fivecmmib='177
\newfam\cmmibfam
\textfont\cmmibfam=\tencmmib \scriptfont\cmmibfam=\sevencmmib
\scriptscriptfont\cmmibfam=\fivecmmib

\def\C{{\sf C}}

\def\bar{\overline}

\def\CP{\sf{CP}}
\def\T{{\sf T}}
\def\P{{\sf P}}

\def\L{{\mathcal L}}

\def\L{{\sf L}}
\def\B{{\sf B}}

\def\be{\begin{equation}}
\def\ee{\end{equation}}
\begin{document}
\begin{titlepage}
\begin{flushright}

\end{flushright}
\vskip 1.5in
\begin{center}
{\bf\Large{Symmetry and Emergence}}

\vskip
0.5cm {Edward Witten} \vskip 0.05in {\small{ \textit{School of
Natural Sciences, Institute for Advanced Study}\vskip -.4cm
{\textit{Einstein Drive, Princeton, NJ 08540 USA}}}
}
\end{center}
\vskip 0.5in
\baselineskip 16pt
\begin{abstract}  I discuss gauge and global symmetries in particle physics, condensed matter physics, and quantum gravity.  In a modern understanding of particle physics,
global symmetries are approximate and gauge symmetries may be emergent.  (Based on a lecture
at the April, 2016 meeting of the American Physical Society in Salt Lake City, Utah.)  \end{abstract}
\date{October, 2017}
\end{titlepage}
\def\Hom{\mathrm{Hom}}

\def\U{{\mathcal U}}

The central role of symmetry was a primary lesson of the physics of the first half of the 20th century.  Accordingly, in the early days of particle
physics, the global symmetries or conservation laws were considered fundamental.
These symmetries included the discrete symmetries of charge conjugation, parity, and time-reversal ($\C$, $\P$, and $\T$), 
and the continuous symmetries associated to conservation of baryon
and lepton number ($\B$ and $\L$).   Later, of course, $\L$ was refined to separate conservation of electron, muon, and tau
 numbers $\L_e$, $\L_\mu$, and $\L_\tau$.

Experiment has shown us that many of these symmetries are only approximate.   In the 1950's, the weak interactions were found to violate $\C$ and $\P$,
and in the 1960's, it turned out that they also violate $\T$.  Much more recently, studies of neutrino oscillations have shown that the lepton number differences
$\L_e-\L_\mu$ and $\L_\mu-\L_\tau$ are not quite conserved.

One can imagine the shock when $\C$, $\P$, and later $\T$ violation were discovered \cite{LY,Wu,CF}.  Why was Nature spoiling perfectly good symmetries?  
And if these symmetries were going to be violated, why were they violated so weakly?

  By the time that violation of the separate lepton number conservation 
laws was discovered, the rise of the Standard Model of particle physics had brought a change in perspective.
To understand this, recall that in the Standard Model, a different kind of symmetry, ``gauge symmetry,'' is primary. Gauge symmetry is familiar in
classical electromagnetism and in General Relativity, and it is central in the Standard Model.  Except for the couplings of
the Higgs particle, the interactions of the Standard Model
are all  determined by gauge symmetry. 

By the time that the Standard Model was written down in the 1960's, it was known that $\C$, $\P$, and $\T$ are not exact symmetries, but baryon and 
lepton number conservation were widely presumed to be fundamental symmetries, though of mysterious origin.  The Standard Model, however, gave a different perspective \cite{Weinberg}.   These symmetries can be interpreted as low energy accidents that are indirect consequences of gauge symmetry.  
The meaning of this statement is that given the gauge symmetries and the  field content (especially the quark
and lepton quantum numbers) in the Standard Model, it is simply
impossible to  find a renormalizable
gauge-invariant operator that violates any of these symmetries at
the classical level.

The operator of lowest dimension that violates lepton number symmetry
 is the dimension 5 operator $HHLL$, where $H=\begin{pmatrix}\phi^+\cr \phi^0\end{pmatrix}$
is the Higgs doublet and $L=\begin{pmatrix}\nu \cr e^-\end{pmatrix}$ is a lepton doublet.   On dimensional grounds, this must be multiplied in the Lagrangian or
the Hamiltonian by a constant with dimensions of inverse mass:
\be\label{conta}{\mathcal L}_1 = \frac{1}{M}HHLL. \ee
After $H$ gets an expectation value, breaking the electroweak gauge symmetry, this interaction leads to
a neutrino Majorana mass  $m_\nu\sim \langle H\rangle^2/M$.   If global symmetries such as $\L_e$, $\L_\mu$, and $\L_\tau$ are supposed to be low 
energy accidents that are indirect results of gauge symmetry, we should expect such a term to be present at some level.  
If we apply the same logic to baryon number, we find in the Standard Model that the operator of lowest dimension that can explicitly violate $\B$ is
a dimension 6 operator
\be\label{contab}{\mathcal L}_2= \frac{1}{M^2}QQQL, \ee
where now $Q$ is a quark multiplet.  

What might we expect $M$ to be?  in the 1970's, physicists tried to guess this based on theories that attempted a ``Grand Unification'' of the particle forces
\cite{Pati,GG}.
The key technical idea here was to use the renormalization group to extrapolate the particle couplings from the energy at which they are measured to a much
higher energy at which the forces can be unified \cite{GQW}.
 From a modern point of view, one might just take as input the observed values of the neutrino mass squared differences.  
Given these values and taking literally the formula $m_\nu\sim \langle H\rangle^2/M$, we find that we need $M$ of roughly $10^{15}\,\mathrm{GeV}$.  
This is beautifully close to the mass scale needed for Grand Unification.  It is also an incredibly high mass scale, much higher
than any fundamental physical mass scale that we can observe in any other way, except via the existence of gravity and possibly
via cosmology.

The observations of neutrino mixing are simultaneously the main
direct support for the existence of new interactions of some kind at a very high energy at which the Standard Model
couplings converge, and also the main support for the idea that the
apparent global symmetries of elementary particles are in significant
part an ``accidental'' consequence of the gauge symmetries of the
Standard Model.  If this interpretation is correct, the proton should decay because of the coupling ${\mathcal L}_2$, and its lifetime might
be close to the experimental bound of about $10^{34}$ years.  

To really clinch this picture, we would like to observe a Majorana mass of the neutrino -- to show that the combined lepton number $\L=\L_e+\L_\mu+\L_\tau$
is violated, and not just the differences of the lepton numbers -- and also observe nucleon decay, to demonstrate violation of $\B$.  
In the
case of the neutrino mass, we have a fairly clear picture of what
sensitivity is needed for a discovery, but this is less so,
unfortunately, in the case of the proton lifetime. But if we could
really observe proton decay, that would be epoch-making and we
would get a lot of new information.

What does this picture say about  $\C$, $\P$, and $\T$?  One basic question is why these symmetries are conserved by the
strong and electromagnetic forces, given that they are not full symmetries of nature.  In the case of electromagnetism, the answer is clear.  Large symmetry violation
would have to be induced by a renormalizable operator, that is one of dimension $\leq 4$; unrenormalizable operators with a  mass scale characteristic of new physics
beyond the strong
and electromagnetic interactions produce small effects,
as above. But there is no
way to perturb Quantum Electrodynamics (QED) by an operator of dimension $\leq 4$ that violates any of its global symmetries, including the ones we have mentioned
and some, such as strangeness, that we have not.
 For the strong interactions or Quantum Chromodynamics (QCD), we {\it almost} get the same answer: with one exception,  QCD does not admit any operator of dimension $\leq 4$ that would
violate any of its observed global symmetries.
 The exception is that $\P$ and $\T$ (and therefore, in view of the $\sf{CPT}$
theorem, also $\CP$ and $\sf{CT}$) can be violated by a ``topological'' coupling
\be\label{gutt}\frac{\theta}{32\pi^2} \,\varepsilon^{\mu\nu\alpha\beta}\mathrm{tr}\,F_{\mu\nu}F_{\alpha\beta}, \ee
where $F_{\mu\nu}$ is the gauge field strength of QCD.  This operator is of dimension 4, so the coupling parameter $\theta$ is dimensionless.
Why  $\theta$ is very close to zero is called the ``strong CP problem.''

A plausible solution, but not yet confirmed experimentally, involves the existence of a very light new particle, the ``axion'' $a$.  The axion 
field\cite{PQ,We,Wi,K,DFS}
is supposed to have an approximate shift symmetry $a\to a+{\mathrm{constant}}$ that is violated primarily by a coupling to QCD of the form
\be\label{utt}{\mathcal L}_3 = \frac{a}{M'} \,\varepsilon^{\mu\nu\alpha\beta}\mathrm{tr}\,F_{\mu\nu}F_{\alpha\beta}. \ee
Given this, the parameter $\theta$ can be eliminated from the low energy physics (to a very high precision) by shifting the value of $a$.  As a result,
the strong interactions will conserve $\P$ and $\T$, as they are observed to do.  Of course, to confirm this picture, one needs to observe
the axion.  Its mass is computable and is of order $m_\pi^2/M'$, where $m_\pi$ is the pion mass.  The axion is a missing link to confirm the idea that ``symmetries are only there to the extent that they are required by gauge symmetry.''

Ignoring the question about the $\theta$ parameter, the status of  strangeness and $\C$, $\P$, and $\T$ and so on in a low
energy world dominated by QCD and QED is comparable to the status of $\L_e-\L_\mu$ or $\L_\mu-\L_\tau$ in the full Standard Model. The symmetries
in question are symmetries of QED and QCD, but they are explicitly broken by dimension 6 operators such as
\be\label{dimsix}\frac{1}{M_W^2}\bar s\gamma_\mu(1-\gamma_5) u \,\bar \nu \gamma^\mu(1-\gamma_5) e. \ee
Historically, observation of such dimension 6 operators pointed to ``new physics'' at what we now know as the weak scale ($M_W$ is now understood as the $W$ boson
mass), rather as neutrino oscillations
plausibly point to some sort of new physics at the traditional scale  of Grand Unification.  

While gauge symmetry makes $\C$, $\P$, and $\T$ automatic in the case of QED and QCD (except for the problem with the $\theta$-angle), it is nearly
the opposite
for weak interactions.  The gauge structure of the weak interactions and the quantum numbers of the quarks and leptons make it impossible for the
weak interactions to conserve $\C$ or $\P$.   This is actually one of the most important insights of the Standard Model.  It prevents the quarks and leptons
from having bare masses and is the reason that there is no analog for fermions of the 
``hierarchy
problem'' concerning the mass of the Higgs particle.    (The hierarchy problem
is the question of what stabilizes the mass of the Higgs particle against potentially very large effects of quantum renormalization.)  The gauge symmetry
of the Standard Model allows the weak interactions to violate $\T$, and it turns out that -- despite the feeble nature of the $\T$ violation that we see
in the real world -- the weak interactions violate $\T$ more or less as much as possible, given the structure of the gauge symmetries and the values of the
quark masses.  

So this is one line of thought that, roughly 40 years ago, led to an expectation that the apparent global symmetries of nature are only approximate.
But in fact, three other lines of thought converged on the same idea in roughly the same period.

First, it turned out that in the Standard Model, though $\B$ and $\L$ are valid symmetries classically, they suffer from a quantum ``anomaly'' and
are not exact symmetries \cite{TH}.  At the time, it was conceivable that the anomaly might be canceled by contributions of yet-unknown fermions and that
$\B$ and $\L$ conservation might be rescued.  By now we know that this is not the case.   (Fermions that are going to contribute to the anomaly
cannot be much heavier than the weak scale, and would have been discovered at the Large Hadron Collider at the CERN laboratory.)  So there is a clear
prediction of $\B$ and $\L$ violation by the Standard Model anomaly, but unfortunately this effect is much too small to be observable, except possibly
in cosmology, and then only under  favorable assumptions.

A second line of thought indicating that gauge symmetries are primary, and global symmetries only approximate, arises from thought experiments
involving black holes.   In 1974, Hawking discovered that black holes evaporate at the quantum level \cite{Hawking}.   In the real world, black holes form from
matter that is rich in baryons and leptons.  But when (in theory) black holes evaporate, we do not get the baryons and leptons back.   So formation
and evaporation of a black hole does not conserve $\B$ or $\L$.  On the other hand, black holes conserve gauge quantum numbers -- such as electric
charge -- because they can be measured by flux integrals at infinity.

This suggests that in a model of Nature complete enough to
include both quantum mechanics and gravity, the only true
symmetries are gauge symmetries. Confirmation comes from the
fact that this turns out to be the situation in String Theory, the
only framework we have for a consistent theory with both quantum
mechanics and gravity. If one looks closely, one always finds that
symmetries in String Theory either are not exact symmetries, or
else they are gauge symmetries.  Sometimes one does have to look closely to see this.

Going back to the black hole, there is an interesting gap in the reasoning.   The thought experiment involving formation and evaporation of a black
hole shows that a theory of quantum gravity cannot have {\it continuous} global symmetries such as the $\mathrm{U}(1)$ symmetry associated to conservation
of $\B$.   But this argument would allow discrete or especially finite symmetry groups such as ${\Bbb Z}_n$, or equivalently it would allow quantities
that are conserved mod $n$ for some integer $n$.   The reason is that we do not understand black hole evaporation nearly well enough to
decide if some mod $n$ conservation law (as opposed to an additive one like baryon number)
might hold in the formation and evaporation of a black hole.

    \begin{figure}
 \begin{center}
   \includegraphics[width=3in]{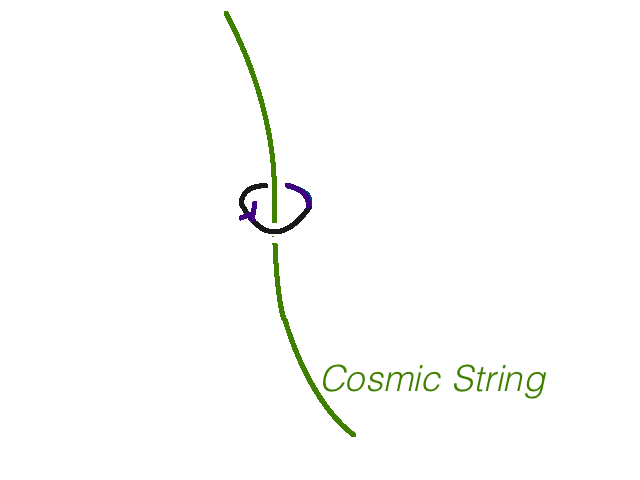}
 \end{center}
\caption{\small  A cosmic string associated to a $\Bbb{Z}_n$ symmetry. \label{FigOne}}
\end{figure}

So in a world with quantum gravity, do we expect discrete global symmetries, or should also discrete symmetries be gauge symmetries?  First we
have to decide what the question means.   A continuous unbroken gauge symmetry is associated to a massless gauge field,
and this is how we distinguish it from a continuous global symmetry; if the symmetry is spontaneously broken, the global symmetry but not the gauge
symmetry leads to the existence of a massless Goldstone boson.  But how do we decide if a {\it discrete} symmetry is a gauge symmetry?

What it means to call a ${\Bbb Z}_n$ symmetry a gauge symmetry is that when one goes around a loop in spacetime, one might come back to the original
state rotated by a symmetry element.  For instance, if a theory has a cosmic string producing a ${\Bbb Z}_n$ rotation (fig. \ref{FigOne}), then this definitely means that the
${\Bbb Z}_n$ symmetry is a gauge symmetry.   With this interpretation of what the question means, the discrete symmetries in String Theory turn out
to be gauge symmetries.  Thus,  in String Theory all of the exact symmetries are gauge symmetries.   This is consistent with what we will
find later when we discuss ``emergence.''

Finally, and also in the period around 1980, the theory of the Inflationary Universe gave a powerful additional hint that $\B$ must not be truly conserved.   Cosmic inflation elegantly explains the near flatness and homogeneity of the Universe.  It has been 
extraordinarily successful at predicting and describing the almost scale-invariant fluctuations in 
the Cosmic Microwave Background (CMB) that are believed to have provided the seeds for galaxy formation.  However, the inflationary universe really only
works if the laws of nature violate $\B$.  The reason for this
is that an early period of exponential expansion of the Universe dilutes the density of matter and radiation to an extremely low level.  Upon the end of
inflation, the Universe can reheat to a reasonable temperature, eventually leading, after further expansion, to the CMB as we see it today.  However, unless
the baryons can be spontaneously generated when (or after) the Universe reheats, we will be left with a world that is symmetrical between matter and
antimatter, very unlike what we observe.   But to spontaneously generate the baryons is only possible if the laws of nature violate $\B$ (and also the discrete symmetries $\C$ and $\CP$ that exchange baryons with antibaryons).

To understand these matters more deeply, we should discuss the physical meaning of gauge and global symmetries.  The meaning of global symmetries
is clear:  they act on physical observables.   Gauge symmetries are more elusive as they typically do not act on physical observables.  Gauge symmetries
are redundancies in the mathematical description of a physical system rather than properties of the system itself.   

One of the important developments in our understanding of Quantum Field Theory that came to fruition in the 1990's (following earlier clues 
\cite{OM}) makes
it clear that this distinction is unavoidable.  Gauge theories that are different classically can turn out
to be equivalent quantum mechanically.  For example, a gauge theory in four spacetime dimensions with gauge group $\mathrm{SO}(2n+1)$ 
and maximal supersymmetry is equivalent to the same theory with gauge group $\mathrm{Sp}(2n)$.   The global symmetry is the same 
in the two descriptions, but the gauge symmetry is different.   It is up to us whether to describe the system using
$\mathrm{SO}(2n+1)$ gauge fields or $\mathrm{Sp}(2n)$ gauge fields.  So neither of the two gauge symmetries is intrinsic to the system.  

Gauge symmetry develops an invariant meaning that must be reflected in any description only if it produces conservation laws  that result from
conserved flux integrals at infinity.  But there are multiple ways for this to fail to happen.  Two such mechanisms
are observed in the Standard Model: the gauge symmetry
of QCD does not lead to conservation laws because of quark confinement, and the gauge symmetry associated to the $W$ and $Z$ bosons of the
weak interactions does not lead
to conservation laws because of spontaneous symmetry breaking.\footnote{A third option, not yet seen in nature, is that gauge symmetry can fail to generate
a conservation law because of infrared divergences that prevent one from defining the would-be conserved quantity.  This is actually what happens
in the example mentioned earlier with $\mathrm{SO}(2n+1)$ or $\mathrm{Sp}(2n)$ gauge symmetry.}
  In the Standard Model with $\mathrm{SU}(3)\times \mathrm{SU}(2)\times \mathrm{U}(1)$
symmetry, only the $\mathrm{U}(1)$ leads to conservation laws (conservation of electric and magnetic charge).

To put it differently, global symmetry is a property of a system, but gauge symmetry in general is a property of a description of a system.  What we really learn
from the centrality of gauge symmetry in modern physics is that physics is described by subtle laws that are ``geometrical.''  This concept is hard to define,
but what it means in practice is that the laws of Nature are subtle in a way that defies efforts to make them explicit without making choices.  The difficulty of
making these laws explicit in a natural and non-redundant way is the reason for ``gauge symmetry.''

We can see the relation between gauge symmetry and global symmetry in another way if we imagine whether physics as we know 
it could one day be derived from something much deeper --  maybe unimaginably deeper than we now have. Maybe the spacetime we 
experience and the particles and fields in it are all ``emergent'' from something much deeper.

Condensed matter physicists are accustomed to such ``emergent'' phenomena, so to get an idea about the status of symmetries in an
emergent description of Nature, we might take a look at what happens in that field.  {\it Global} symmetries that emerge in a low
energy limit are commonplace in condensed matter physics.   But they are always {\it approximate} symmetries that are
explicitly violated by operators of higher dimension that are ``irrelevant'' in the renormalization group sense.   Thus the global symmetries in
emergent descriptions of condensed matter systems are always analogous to $\L_e-\L_\mu$ or $\L_\mu-\L_\tau$ in the Standard Model -- or to
strangeness,  etc., from the point of view of QED or QCD.

By contrast, useful low energy descriptions of condensed matter systems can often have exact gauge symmetries that are ``emergent,'' meaning that
they do not have any particular meaning in the microscopic  Schrodinger equation for electrons and
nuclei.  The most familiar example would be the emergent $\mathrm{U}(1)$ gauge symmetries that are often used in effective field theories of the fractional
quantum Hall effect in $2+1$ dimensions.  These are indeed {\it exact} gauge symmetries, not explicitly broken by high dimension operators. 
Gauge theory with explicit gauge symmetry breaking is not ordinarily a useful concept. 

An emergent gauge theory in condensed matter physics is never a ``pure gauge theory'' without charged fields.  On the contrary, such a theory
always has quasiparticles from whose charges one can make all possible representations of $G$.   Otherwise, from the effective theory of the emergent
gauge field, one could deduce exact degeneracies among energy levels that have no natural interpretation in the underlying Schrodinger equation of
electrons and nuclei.  For the same reason, an
 emergent gauge theory in condensed matter physics will contain all of the magnetic objects whose existence
is suggested by the low energy physics; the details depend on $G$ and on the spacetime dimension.  For $G={\mathrm{U}}(1)$,  the magnetic
objects are instantons in $2+1$ dimensions (corresponding in condensed matter physics to a thin film) and magnetic monopoles  in $3+1$ dimensions. 
For $G$ a finite group, there are  vortex quasiparticles in $2+1$ dimensions and strings in $3+1$ dimensions, as sketched in fig. \ref{FigOne}.

This has an echo in quantum gravity -- or at least in String Theory, where we are able to test the matter. In String Theory, gauge fields always couple to the full
complement of electric and magnetic charges suggested by the low energy description.   This depends ultimately on a rather
subtle calculation \cite{Polchinski}.

    \begin{figure}
 \begin{center}
   \includegraphics[width=4in]{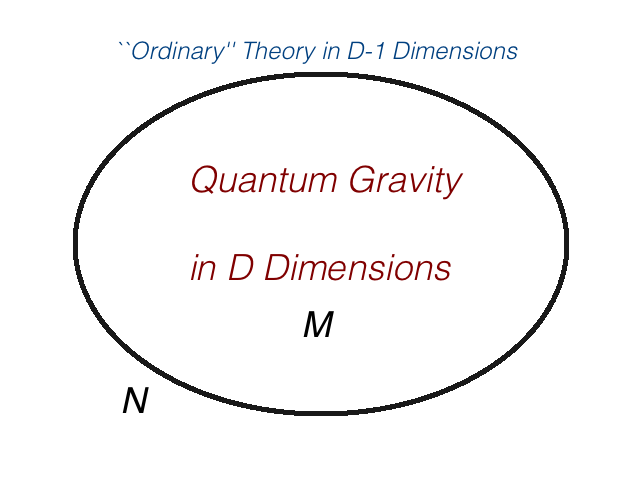}
 \end{center}
\caption{\small  Duality between quantum gravity in a $D$-dimensional spacetime $M$ and an ``ordinary theory,'' which here just means a quantum
field theory without gravity, on the conformal boundary $N$ of $M$. \label{FigTwo}}
\end{figure}

In the context of quantum gravity, we actually do have an interesting and informative framework for an ``emergent'' description of something like
the real world -- or at least of a world with quantum gravity together with other particles and forces.  
This is the gauge/gravity or AdS/CFT  duality.\footnote{AdS
stands for Anti de Sitter spacetime, the analog of Minkowski spacetime with negative cosmological constant.  CFT is conformal field theory.  In the simplest
examples of gauge/gravity duality, the quantum gravity propagates in an asymptotically AdS spacetime, and the gauge theory is a CFT.}  Here
the spacetime with its gravitational metric and all the fields in it are ``emergent'' from a description by an ``ordinary theory'' on the conformal boundary of
spacetime (fig. \ref{FigTwo}). In this context, an ``ordinary theory'' is just a quantum field theory without gravity (typically but not necessarily
a gauge theory).   Gauge/gravity or AdS/CFT 
duality can be described in an abstract way, but the concrete examples in which we know something about each side
of the duality come from String Theory.

In gauge/gravity or AdS/CFT duality, one starts with an ordinary theory on a spacetime $N$ of some dimension $D-1$.  The gravitational dual
is formulated on $D$-dimensional spacetimes $M$ that have $N$ for their conformal boundary. (This means roughly that $N$ lies at infinity on $M$.)
 In general, given $N$, there is no distinguished $M$, and one
has to allow contributions of all possible $M$'s.   This is as one should expect: in quantum gravity, spacetime is free to fluctuate, and this includes
the possibility of a fluctuation in the topology of spacetime.  Only the asymptotic behavior of spacetime -- here the choice of $N$ -- is kept fixed
while the spacetime fluctuates. 

Now suppose that the theory on $N$ has a global
symmetry group $G$.  Then one can couple the theory on $N$ to a background classical gauge field\footnote{One should ask whether there is an 't Hooft anomaly
obstructing this coupling.  In practice, the only such anomaly is a $c$-number.  The reasoning given in the text remains valid in the presence
of such a $c$-number 't Hooft anomaly.  The $c$-number 't Hooft anomaly on $N$ corresponds to a Chern-Simons coupling for the gauge fields on $M$.} $A$ with that
gauge group.  In this situation, the statement of the duality involves
an extension of $A$ over $M$.  But just as there was no natural way to pick $M$, there is no natural way to pick the extension of $A$ over $M$.
So just as we 
 have to sum over the choice of $M$, we have to sum or integrate
over all possible extensions of $A$ over $M$.  But summing or integrating over the extension of $A$ over $M$ means that $A$ is a quantum gauge field on $M$
(whose boundary value on $N$ is fixed).  So if there is a global symmetry $G$ on $N$, then the dual theory has a quantum gauge symmetry $G$ on $M$.  Note that
this reasoning applies equally whether $G$ is a continuous group like $\mathrm{U}(1)$ or a finite group like $\Bbb{Z}_n$.

By contrast, if the theory on $N$ (in some way of describing it) has a gauge symmetry, this does not correspond to anything simple on $M$.   The theory on $M$
has gauge symmetries, which correspond to global symmetries on $N$, but it does not have global symmetries.  

In trying to loosely extrapolate the gauge/gravity
duality to the real world, we ourselves correspond to observers on $M$ (since we experience gravity) so we would see gauge symmetries but not exact global symmetries. The most
general lesson of the known gauge/gravity duality is that the ``ordinary theory'' from which gravity emerges is formulated not on $M$ but on another space $N$.  ``Emergence''
means the emergence not just of the gravitational field but of the spacetime $M$ on which the gravitational field propagates.  Any emergent theory of gravity will have this
property, since an essential part of gravity is that $M$ is free to fluctuate and cannot be built in from the beginning.

Going back to particle physics, 
it is striking how the modern understanding of
symmetries in particle physics is consistent with the idea that the spacetime we live in 
and all the particles and forces in it are emergent in a way somewhat similar to what  happens in gauge/gravity or AdS/CFT duality.   
This interpretation of the world implies that there should be no true global symmetries in nature, so  the violation of  $\L_e-\L_\mu$ and $\L_\mu-\L_\tau$ that has been observed in neutrino oscillations removes a potential obstacle.  Of course, matters would become clearer if we could also observe the Majorana mass of the
neutrino and the decay of the proton -- and for good measure if we could find a QCD axion.  In fact, if an axion is discovered, its coupling
to QCD would itself give an example -- like others we have discussed -- of an approximate global
symmetry that is explicitly broken by an operator of higher dimension.  In this case, the symmetry is the shift symmetry of the axion, and the dimension 5
operator that breaks the symmetry was written in eqn. (\ref{utt}).

Research supported in part by NSF Grant PHY-1606531.

\bibliographystyle{unsrt}

\end{document}